
\input phyzzx

\def\dfdg{{\partial F\over \partial G}}
\def\dfdrh{{\partial F\over \partial r_h}}
\def\pr{p_{\hat r}}
\def\pph{\left .(\rho +\pr )\right |_{r_H }}
\def\pgr{(\phi  ;G,r_H )}
\def\pbgr{(\bar{\phi}  ;G,r_H )}
\def\pb{ {\bar \phi}}
\def\dedp{{{\delta E}\over {\delta \phi }}}
\def\lm{L_m(\phi)}
\def\pbi{{\bar \phi}_i}
\def\pbio{{\bar \phi}_{io}}
\def\ppri{\phi^\prime}

\REF\coleman{S.Coleman, Nucl. Phys.{\bf B262}, 263 (1985).}
\REF\was{M.Colpi, S.Shapiro and I.Wasserman,Phys.Rev.Let.{\bf 57}, 2485(1986).}
\REF\moss{Lucock and I. Moss, Phys. Lett. {\bf 176B}, 341 (1986).}
\REF\glen{N.K. Glendenning, T. Kodama and F.R. Klinkhamer, Phys. Rev.
{\bf D38}, 3226 (1988).}
\REF\droz{S. Droz, M. Heusler and N. Straumann, Phys. Lett. {\bf 268B},
371 (1991); {\bf 271}, 61 (1991).}
\REF\bizon{P. Bizon and T. Chmaj, {\it Gravitating Skyrmions}, preprint
UWThPh-1992-23 (1992).}
\REF\weinberg{K. Lee, V.P. Nair, and E.J. Weinberg, Phys. Rev. {\bf D45},
2751 (1992).}
\REF\reissner{F.A. Bais and R.J. Russell, Phys. Rev. {\bf D11}, 2692
(1975); Y.M. Cho and P.G.O. Freund, Phys. Rev. {\bf D12}, 1588 (1975).}
\REF\kylee{K. Lee, V.P. Nair and E.J. Weinberg, Phys. Rev. Lett.
{\bf 68}, 1100 (1992).}
\REF\qstarsa{B.W. Lynn, Nucl. Phys. {\bf B321}, 465 (1989).}
\REF\qstarsb{S. Bachall, B.W. Lynn and S.B. Selipsky, Nucl. Phys. {\bf B331};
67 (1990); {\bf B225}, 606 (1989), {\bf B321}, 465 (1989).}
\REF\lid{A.Liddle and M.Madsen, {\it The Structure and Formation of Boson
Stars }, preprint SUSSEX-AST 92/2 (1992).}
\REF\strange{see e.g. A. Olinto, {\it The Physics of Strange Matter}, preprint
FERMILAB-Conf-91/349-A (1991)}
\REF\vn{P. van Nieuwenhuizen, D. Wilkinson, and M.J. Perry, Phys. Rev.
{\bf D13}, 778 (1976).}
\REF\choquet{see for example, Y. Choquet-Bruhat, C. DeWitt-Morrette and
M. Dillard-Bleick, {\it Analysis, Manifolds, and Physics}, North-Holland
(1982).}
\REF\vilenkin{M. Barriola and A. Vilenkin, Phys. Rev. Lett. {\bf 63}, 341
(1989)}
\REF\nohair{S. Adler and R. Pearson, Phys. Rev. {\bf D18}, 2798 (1978).}
\REF\visser{M. Visser, {\it Dirty Black Holes: Thermodynamics and Horizon
Structure}, preprint WASH-U-HEP-91-71-REV (1992).}
\REF\bekenstein{J.D. Bekenstein, Phys. Rev. {\bf D5}, 1239 (1972).}

\def\ackn{\foot{This work was supported in part by NSF grant
NSF-THY-8714-684-A01}}
\Pubnum={UMHEP-374}
\date{July, 1992}
\titlepage
\title{Horizons Inside Classical Lumps\ackn}
\author{David Kastor and Jennie Traschen}
\address{Department of Physics and Astronomy\break
         University of Massachusetts\break
         Amherst, Massachusetts 01003}
\vfil
\abstract
We investigate the possibility of having horizons inside various
classical field configurations.
Using the implicit function theorem, we show that models satisfying a
certain set of criteria allow for (at least) small
horizons within extended matter fields.  Gauge and global monopoles
and Skyrmions satisfy these criteria.  Q-balls and Boson stars are
examples which do not and can be shown not to allow for horizons.
In examples that do allow for horizons, we show how standard
`no hair' arguments are avoided.

\endpage

\chapter{Introduction}
Black holes are intriguing objects and worth studying in all
their possible varieties.
In this paper we will study
the possibility of having black holes
inside various classical field configurations.
Examples we consider include gauge and global monopoles, Skyrmions,
Q-balls [\coleman], and Boson stars [\was].

Besides the basic search for black hole
solutions, there are a number of physically motivated questions one
can ask in this context.  For instance, what happens when you drop
such an object into a Schwarschild black hole?  For a gauge
or global monopole the result should be a black hole with
the appropriate kind of hair, since these both involve non-trivial
behavior of the fields at infinity.  But in the case of a Skyrmion,
one might think that the only possibility would be its vanishing
without a trace. Our results show that there is another
possibility, at least for horizons very
small compared to the Skyrmion radius\foot{numerical results on extended
Skyrmion fields around a black hole are given in
references [\moss,\glen,\droz,\bizon].}.
In the case of gauge monopoles,
Lee et.al. [\weinberg] have argued that, besides the Reissner-Nordstrom
type solutions [\reissner], there also exist,
for sufficiently small horizon radius, solutions in which the Higgs field
and gauge field behave more like an extended monopole outside the horizon.
Our results confirm their arguments, and show that global monopoles
can also have horizons inside them.

Horizons inside extended field configurations may also be relevant in the
late stages of black hole evaporation by Hawking radiation.  Lee et.al.
[\kylee] have shown that extreme, magnetic Reissner-Nordstrom type black
holes are unstable in a theory with extended monopole solutions.  They
conjecture that the extended solution discussed above is stable and that
evaporation of the black hole proceeds through this configuration, leaving
a non-singular magnetic monopole as the end state.  Perhaps a Schwarschild
black hole in a Skyrmion theory, for example, similarly becomes unstable (or
metastable) when its radius is less than the characteristic Skyrmion radius.
The evaporation process may then leave behind other stable remnants.

Finally, in the literature Q-stars (large Q-balls) [\qstarsa,\qstarsb]
and Boson stars (see [\lid] and references therein), as well as strange
matter [\strange] and other types of non-topological solitons, are discussed as
candidates
for compact astrophysical objects.  We can ask what the possible final
collapsed states of such matter are.

\chapter{Existence of Solutions with Horizons}
We will be looking for static, spherically symmetric solutions to
Einstein's equation, which have nonsingular, nontrivial matter fields
outside a horizon.  The form of the metric
will be taken to be
$$ds^2 =-B(r) dt^2 +A(r)dr^2 +r^2 d\Omega ^2 \eqn\metric$$
It is often convenient to define the function $m(r)$ by
$${1\over {A(r)}} =1-{{2Gm(r)}\over r} .\eqn\ainv$$
A horizon occurs at coordinate $ r_H$ if
$$2Gm(r_H )= r_H .\eqn\horizon$$
When a horizon is present, one also
expects that $m_o\equiv m(0) \not= 0$, so that the metric is
not well behaved at the origin. This is like having a seed mass at the
origin.

Let us agree to call a {\it star} a configuration of matter fields
$\phi $ (not necessarily scalar)
such that the stress-energy is static, spherically symmetric,
and localized.
Suppose a particlar matter
field theory has star type solutions, without gravity. There is
some force balance, without gravity, which keeps the field configuration
from either
collapsing to a point or expanding to infinity. One might expect that
weakly gravitating solutions would then exist, and that even placing a
small seed mass inside the star, wouldn't disturb the balance too much.
This can be made more precise by considering the
Oppenheimer-Volkoff (OV)
equation of hydrostatic equilibrium, which states (in the case when
the three principal pressures are not necessarily the same)
$${{d\pr}\over {dr}}= - G{(m(r) +4\pi r^3 \pr )\over r(r-2Gm)}
(\rho +\pr   )  + {2\over r}(p_{\hat{ \phi}} -\pr )\eqn\ov $$
In the absence of gravity, only the second term on the right hand side
is present and for weak gravity, this term may still dominate.
However, from the first term in \ov , we see that at a horizon
the sum of the radial pressure and the
energy density must vanish.
In a normal, burning star, both these quantities are positive
and a horizon is not possible.  On the other hand
for many field theories, it happens quite naturally that $\pph =0$.

Our main result will be to show that given a matter theory
which ($1$) has star type solutions without gravity and ($2$) satisfies
$\pph =0$ ``automatically'' (in a sense defined below),
then there exist star
solutions when the matter theory
is coupled weakly to gravity, and there also exist solutions
with horizons inside.

More precisely, the non-gravitating matter theory is
described by a Lagrangian $L_m$.  A star solution is found
by evaluating the action on field configurations consistent with a particular
static, spherically symmetric ansatz.  The Lagrangian restricted to this class
of fields will be written $\lm$.  We will assume that
$-\lm $ is positive definite.  When the matter theory is coupled to gravity,
we will
assume that the sum of the energy density and radial pressure is given by
$$(\pr +\rho )={1\over A}K(\phi ),\eqn\prho$$
Where $K$ is a functional of the matter fields only. Then there exist
regular star type solutions to the Einstein equation, and there also exist
star type solutions horizons, which have nontrivial, nonsingular
matter fields outside the horizon, for $G$ and $r_H$ sufficiently small.
The argument, as follows, is an application of the implicit function
theorem.

First define new gravitational variables,
$$e^x =\sqrt{{B\over A}}\qquad{\rm and} \qquad e^y =\sqrt{AB}.\eqn\newvar$$
The action for fields outside a
horizon is then taken to be
$S= \tilde S _E  +S_m$, with
$$\eqalign{\tilde S _E (x,y)&=-{1\over 8\pi G}\int _{r_H} ^{\infty}
dr y^{\prime} ((r-r_H )e^y -re^x )\cr
S_m (\phi _n ,x, y)&=\int _{r_H} ^{\infty}
dr r^2 e^y L_m\cr}\eqn\action $$
$\tilde{S}_E$ differs from the usual Einstein action by a boundary term, which
has been chosen so that varying $\tilde{S}_E$ imposes the correct boundary
condition at the horizon (see reference [\vn]).
Varying the action with respect
to $x$ and $y$ gives the equations of motion
$$y' = -8\pi G r e^{y-x}{{\delta L_m }\over {\delta x}}\eqn\ymotion$$
$${d\over {dr}}(r(e^y -e^x ))= -r^2 e^y (e^{y-x} {{\delta L_m }\over
{\delta x}} +L_m + {{\delta L_m }\over {\delta y}})\eqn\xmotion$$
and the boundary condition
$$\left . re^x \right |_{r_H} = r\sqrt{{B\over A}}\left . \right |_{r_H}
=0.\eqn\bound$$
Note that from the definition of the stress tensor
$-2{{\delta L_m}\over {\delta x}}= \pr +\rho $.
Equations \ymotion\ and \xmotion\ can be used to solve for
the gravitational fields $x$
and $y$ in terms of the matter fields alone if and
only if $\pr +\rho ={1\over A}K(\phi )$, where $K$ is a function of
the matter fields alone.  This was one of our
assumptions.  This is equivalent to
the matter lagrangian having the form
$$\lm =-{1\over A}K(\phi )-U(\phi ,AB)\eqn\matter$$
We can then define a positive definite functional of the fields,
$E(\phi,G,r_H)$, by
$$E\pgr =-S =\int _{r_H} ^{\infty} dr e^y (r(r-r_H )K +r^2 U)\eqn\energy$$
In \energy , $y(r)$ is given in terms of the matter fields by
$$y(r)= -8\pi G\int _r ^{\infty} dr' r' K(\phi)\eqn\solve$$
Note that for a given configuration of the fields $\phi$,
$E(\phi,G,r_H)$ is a continuous, differentiable function of $G$ and $r_H$.

We assume that for $G=r_H =0$, the functional $E(\phi,0,0)$
has a minimum $\pb _{o}$. This is our non-gravitating star.
For $G$ and $r_H$ nonzero,
we seek solutions $\pb  $ to
$$F(\pbgr )\equiv \dedp =0,\eqn\motion,$$
which by construction will satisfy the equation of motion with the correct
boundary conditions.
By assumption $F(\pb _{o} ;0,0)=0$.
The implicit function theorem for Banach spaces
\foot{In the appendix we sketch a finite dimensional
version of the theorem, which illustrates the relevant points.}
[\choquet] can then be used to show
that for $G$ and $r_H$ sufficiently close to
zero, there exist functions $\pb (G,r_h )$ satisfying \motion ,
such that
$\pb  (0,0)=\pb _{o}$.
This can be seen by expanding \motion ,
$$ 0= {\delta F\over \delta\phi}\cdot(\pb-\pb_0)+
{\partial F\over \partial G}\cdot G +
{\partial F\over \partial r_H}\cdot r_h +\dots, \eqn\expand $$
with all the derivatives evaluated at $\phi=\pb_0$, $G=0$, and
$r_H=0$.
There will be a solution for $\pb$ as long as the operator
${\delta  F\over\delta\phi}$ in \expand\ is an isomorphism between two
Banach spaces $H_1$ and $H_2$, and the two functions $\dfdg$ and $\dfdrh$
belong
to the space $H_2$.  The choice of particular function spaces depends on the
system under consideration.  However, roughly speaking, we can see that this
will be true in general given that the flat space solution
$\pb_0$ is a minimum of the
energy functional \energy\ , which is equivalent to
$$\left .{\delta F\over\delta \phi}
\right | _{(\pb_0,0,0)}\cdot \delta \phi> 0.\eqn\minimum$$
Hence ${\delta F \over \delta \phi}$ has no zero modes and is invertable.
In the next section
we indicate how to choose appropriate function spaces for global
monopoles.

The OV equation implied that $(\pr +
\rho )\propto {1\over A}$ at a horizon.
Above, we found that this same condition
was needed to integrate out the metric coefficients $A$ and $B$ from the
action.  This allowed us
to use the existence of non-gravitating solutions to imply via
the implicit function theorem the existence of gravitating solutions
and solutions with horizons.
If we take a theory, such as Q-balls, in which, as we will see below,
$A$ and $B$
cannot be eliminated from the action, then
to use the implicit
function theorem, one would have to compute the variation including all
the dependent functions, $\phi,A$ and $B$. But knowledge of
the flat space solutions gives us no information analogous to \minimum\
about variations
in the $A$ or $B$ directions, so the argument can't proceed.



\chapter{Global Monopoles}

In this section we demonstrate the use of the implicit function
theorem and selection of appropriate function spaces for global
monopoles. The matter field theory for the basic
global monopole is given by an
$SO(3)$ invariant Lagrangian for a triplet of scalar fields $\phi^a$,
$${\cal L}= \half \nabla^\mu\phi^a \nabla_\mu\phi_a - \half\lambda
\left (\phi^a\phi_a-v^2\right )^2,\eqn\global$$
where $\nabla_\mu$ is the covariant derivative operator.
The scalar field configuration for the monopole has the spherically
symmetric form
$$ \phi^a = v\phi (r)\hat{r}^a.  \eqn\ansatze$$
For solutions without horizons $\phi (r)$ interpolates between $0$ at the
origin and $1$ at infinity.
Evaluated on such field configurations (with the covariant derivative
operator appropriate for the spherically symmetric metric \metric)
the lagrangian has the form
$L_m ={1\over A}K +U$, where the kinetic and potential terms are
given by
$$K=  {1\over 2}v^2 \phi^{\prime 2} ,\qquad
U= {{v^2 \phi^2}\over {r^2}} +{1\over 2}\lambda v^4 (\phi^2 -1)^2,
\eqn\globaltwo $$
Here $\phi^\prime=d\phi /dr$.
The equations of motion for the metric coefficients are
$$m^\prime (r)=4\pi r^2 ({1\over A}K+U),\qquad
{{(AB)^\prime}\over {(AB)}} =16\pi G rK.\eqn\metriceqns $$

The flat space global monopole solution has the following asymptotic
behavior
$$ \bar{\phi}_0(r) \sim \cases{ar,&$r\rightarrow 0$;\cr
                    1-{1\over 2\lambda v^2 r^2},
                     &$r\rightarrow\infty$,\cr}  \eqn\limits$$
where $a$ and $b$ are constants (the
slope $a$ at the origin must be determined numerically).
{}From \limits\ and \globaltwo , one can see that the energy density for the
global monopole falls off only as $1/r^2$, so that the total energy of a global
monopole diverges,
$$\lim_{r\to\infty} m(r)= 4\pi v^2 r. \eqn\rtoinfty$$
Hence the spacetime of a
global monopole is not asymptotic to flat spacetime, but rather to
flat spacetime minus a missing solid angle [\vilenkin],
$$\lim_{r\to\infty}{1\over A} = 1-8\pi G v^2 .\eqn\solid$$
In order to avoid a horizon at large
radius (which is not of the sort we are interested in), we will keep
$8\pi G v^2 <1$.

The quantity ${\delta F\over \delta \phi}$ in \expand\ for the global
monopole is given by
$$ {\delta F\over \delta \phi} \delta\phi = - {d\over dr}\left (
r^2 {d\over dr} \delta\phi\right ) + \left (2 + r^2 \left [
6 \bar{\phi}^2-2\right]\right)\delta\phi \eqn\dfdphi $$
Here we have rescale lengths by a factor $\sqrt{\lambda v^2}$.
The variations $\dfdg$ and $\dfdrh$ evaluated on the background solution can
be seen to have the forms
$$\dfdg \sim \cases{r,&$r\rightarrow 0$;\cr
                    {1\over r^2},&$r\rightarrow\infty$,\cr} \qquad
\dfdrh \sim \cases{const,&$r\rightarrow 0$;\cr
                    {1\over r^3},&$r\rightarrow\infty$.\cr}
       \eqn\vary$$
If we take the variation $\delta\phi$ to have the assymptotic behavior
$$\delta\phi \sim \cases{const,&$r\rightarrow 0$;\cr
                    {1\over r^4},&$r\rightarrow\infty$,\cr} \eqn\pert $$
(with the standard $L^2$ norm in three dimensions),
then we can accomodate the variations induced by
\vary .  This can be seen by examining the asymptotic
behavior of ${\delta F\over\delta\phi}$ in \dfdphi .
We then have to show that the operator
$L={\delta F\over \delta\phi}$ is an isomorphism
between these spaces.  Since the operator
is elliptic, this will be the case if neither it
nor its adjoint have zero modes.
Suppose that $L$ has a zero mode, then we can write
$$ 0 = \int_0^\infty dr\left\{- f {d\over dr} \left (r^2 {d\over dr} f\right )
+r^2 {\delta ^2 U\over \delta\phi ^2} f^2\right\}. \eqn\integral $$
Integration by parts yields
$$ 0 =- r^2 \left . f {d\over dr} f \right |_0^\infty
+ \int_0^\infty dr r^2 \left\{ ({d\over dr}f)^2
+ {\delta^2 U\over \delta\phi ^2} f^2 \right\}. \eqn\contra $$
The boundary term vanishes for functions $f$ having the behavior \pert .
Equation \contra\ then leads to a contradiction if
$$\left . {\delta^2 U\over \delta\phi ^2}\right | _{(\bar{\phi}_0,0,0)}
\geq 0\eqn\positive $$
holds everywhere.  We have checked numerically that \positive\ is satisfied for
the flat space monopole.  Therefore the operator $L$, which is self-adjoint has
no zero-modes.

\chapter{Examples}

Three examples of field configurations which allow horizons inside
are Skyrmions, gauge monopoles, and global monopoles.
These three examples span a range of types:
gauge monopoles have both a long range
magnetic field and topological winding, global monopoles have
only the topological constraint, and the Skyrmion field winds
but is not topological.
These all have $\lm$ of the form \matter , and so satisfy the condition
$\pph=0$ at a horizon.
The implicit function theorem argument
shows that solutions with hair exist for $G$ and $r_H$ in some
range about zero,
but gives no information about how large this range is. One can deduce
more information about the range from arguments based on the traditional
positive `no-hair' integrals, which we do below in Section $5$.

Field configurations which cannot support horizons
include Q-Balls [\coleman] and boson stars [\was].
Q-Balls are star type configurations that exist
without gravity [\coleman], but, as we will see,
fail to satisy the condition $\pph = 0$
at a horizon.
The simplest Q-balls occur in the theory of a single complex scalar
field [\coleman].  The Q-ball field has the form
$\phi =f(r) e^{-i\omega t}$ where $f(r)$ vanishes at infinity.
The lagrangian
evaluated on such configurations is
$$L_Q = {1\over {2A}}(f')^2 +{1\over 2}(m^2 -{{\omega ^2}\over B} f^2)
+U (f^2 ), \eqn\qlagrangian $$
where the mass-term in the potential has been separated out. The frequency
$\omega$  must satisfy $\omega^2 > m^2$ for stability. From the
definition of the stress tensor we then have
$$\pr +\rho =-{2\over A}{{\delta L_m}\over
{\delta 1/A}}+{2\over B}{{\delta L_m}\over {\delta 1/B}}
= -{1\over A}(f')^2 -{1\over B}\omega ^2 f^2.\eqn\qballs $$
We see that to satisfy $\pph=0$, $f$ must vanish at a horizon
\foot{We assume that the volume element $\sqrt{AB}$
is well behaved at a horizon,
which implies that $B\sim r-r_H $ near the horizon.}.
But this means
that the field is in its vacuum both at the horizon and at infinity,
which is not a Q-Ball type solution.

Boson stars (see [\lid] for a review)
are localised scalar field configurations which exist {\it only} with gravity.
The matter lagrangian again has the form \qballs\ (with different
potential terms and with $\omega^2 > m^2$).
Hence Boson
stars satisfy $\pph=0$ only for $f(r_H)=0$, implying again that the
field be in its vacuum at the horizon, as well as at infinity.

A third example which probably does not allow hair is the
Abelian-Higgs model [\nohair].
If the scalar field has the form
$f(r)$ and  the gauge field is given by $A_t (r)$,
then the matter lagrangian is
$$L_{AH}= {1\over {2A}}(f')^2 -{1\over {AB}}(A_t ')^2 -
{1\over {2B}}e^2 (A_t )^2 f^2 +{{\lambda}\over 2}(f^2 -v^2 )^2 \eqn\abelhiggs$$
This again is not of the form \matter , and satisfying
$\pph=0$ requires that
$A_t ^2 f^2 =0$ at $r=r_H$. While this in itself is not enough to rule
out solutions, it clearly makes it ``harder'' to satisfy the equations
given this additional condition on the fields. Indeed, the `no-hair' integrals
discussed in section $5$ further imply that if $A_t (r_H )=0$, then the fields
are in their vacuum states everywhere outside the horizon. Adler and
Pearson [\nohair] explicitly analyzed
the Einstein equation for this system further, and
have shown that this is indeed the case.

Finally, it is interesting to think about the case of a Coulombic electric
field due to a point charge. This is outside the framework of
the present discussion, because the non-gravitating configurations
are singular, $A_t =q/r $. However, the Reissner-Nordstrom
charged black holes $are$ solutions with nonzero,
nonvacuum, regular matter fields outside the horizon
\foot{Visser [\visser] has independently studied the condition $\pph=0$
in the context of various recent
black hole solutions in field theories, such as
dilatons and axions, coupled to gravity.  He has also looked at the
thermodynamics of such solutions.}. In this case,
it is easy to check that the E\&M Lagrangian reduces to
$$L_{EM} ={1\over {AB}}(A_t ')^2\eqn\em$$
which has the form \matter\ and that, in fact,
the combination $\pr +\rho $ vanishes everywhere.

In looking at these various examples, one notices that different kinds
of mass terms play quite different roles. A ``true'' mass, or any potential
$U$ which is independent of the metric, makes no contribution to
the sum $\pr +\rho$, as in Inflation. A dynamical mass which comes
from the coupling to the time component of a gauge potential, contributes
a term to $\pr +\rho \propto {1\over B}f^2 A_t ^2 $, which tends to
rule out hair. A dynamical mass which comes from coupling to
the spatial components of a gauge field contributes zero, and contributes
a winding term $\propto {1\over r^2}$ to $p_{\hat{\phi}}-\pr$, which
is important in the OV equation \ov .

\chapter{`No-Hair' Integrals}

It is interesting to see how the black hole solutions discussed above
avoid being ruled out by standard `no-hair' arguments.  In the case of
extended gauge monopole solutions, this was discussed in ref. [\weinberg].
We will see that Skyrmions and global monopoles escape in basically the
same way.
Necessary conditions for the existence of black hole
solutions in a given field theory can be derived by constructing
energy integrals from the equations of motion
(see e.g. [\nohair,\bekenstein]). If the action in the region
outside the horizon is given by
$$S=-\int _{r_H} ^{\infty} dr J(r),\eqn\genaction $$
an extremum occurs when
$${d\over {dr}}{{\delta J}\over {\delta\phi  '}}={{\delta J}\over
{\delta \phi }},\eqn\euler$$
with the boundary conditions $\delta J/
\delta \phi  '=0$ at $r=r_H$ and the fields going to their vacuum
values at infinity.  Therefore
$$\int _{r_H} ^{\infty}dr \left [ \ppri {\delta J\over \delta \ppri}
+(\phi -\phi_\infty){\delta J\over \delta \phi}\right ]
 = (\phi -\phi_\infty)\left .
{\delta J\over\delta\phi^\prime}\right |_{r_H}=0\eqn\hair $$
Consider the case at hand \energy , where $S=-E$ and $J$ is the positive
definite integrand. Since we are assuming that regular solutions
exist when $G=r_H =0$, the above is true with $r_h =0$ and $e^y \equiv
1$ in $J$. Since typically the gradient term in the integrand
is of the form $C^2 (\phi )(\phi ')^2 $,
this requires that as $r$ ranges from zero to $\infty$, there are
positive and negative contributions to the potential (the second)
term in the integrand. Now, if the lower limit is taken to be
$r_H$, there is still a possibility for positive and
negative contributions
to sum to zero above, if $r_H$ is small enough. This point was discussed
in [\weinberg] in reference to gauge monopoles, noting that the fields had
to be Reissner-Nordstrom outside the horizon if $r_H$ were sufficiently
large. For Skyrmions, the structure of the no-hair integrals depends
on what the response is of the Skyrmion field to gravity. But assuming
that the effect of gravity is to further concentrate the energy
density, again there will be a critical value of $r_H$, such that
if the horizon is larger, the field must be in its vacuum outside
the horizon. On the other hand, global monopoles have
no such restriction on the value of $r_H$.

\ack
We would like to thank Karen Uhlenbeck for helpful and informative discussions
and the Aspen Center for Physics for its hospitality during part of this work.

\Appendix{A}

Here we recall the arguement for the implicit function theorem for
a system of $N$ equations in $N$ unknowns, and the limit as $N$ becomes
a continuous variable. Let $g$ be the independent variable, and
$\pi \ , i=1,...,N$ be $N$ dependent variables. (These are numbers, not
functions.) We seek solutions $\pi =\pbi (g)$ to the system
$$F_j (\pi ,g)=0 \  , j=1,...,N \  ,\eqn\finite $$
given that $\pbio$ is a solution when $g=0$, $F( \pbio ,0)=0$.
Let $\pi -\pbio =\delta \pi $ and denote the matrix of first
derivatives with respect to the independent variables by
$O_{ji} =-{\partial F_j\over \partial\pi}$, evaluated at $\pbio ,g=0$.
Then Taylor expanding the equation $F=0$, to linear order one needs to
solve
$$ O_{ji}\delta \pi =- {\partial F_j \over \partial g}\cdot g \eqn\taylor$$
There is a solution $\delta \pi$ for any ``source'' on the right hand
side of (A.2) if the matrix $O_{ij}$ has no zero eigenvectors, i.e.,
$$O_{ij}v^i v^j \not= 0 \ , {\rm for\   all}\  v^i \eqn\posdef$$

For an implicit functional theorem, we would like the limit where
the discrete index $i$ becomes a continuous variable $x$, with
$F_j \rightarrow F(x) ,\  \pi \rightarrow \phi (x) $. Let $\{ P_i (x)\}$
be a set of basis functions, and let $\phi (x)=\Sigma _i A_i P_i (x)$ and
$\delta\phi (x)=\Sigma _i \delta A_i P_i (x)$. Then in this limit,
$$\Sigma _i {\partial F_j \over \partial \pi}\delta \pi \rightarrow
\int dy {\delta F(x)\over \delta\phi (y)}\delta \phi (y) =\
\Sigma _i {\delta F\over \delta A_i}\delta A_i .\eqn\continuous $$
Hence for a solution
one needs that this last quantity, evaluated at the known solution,
has no zero modes. In the main part of the paper, this condition
was met since the second variation of the energy functional was
nonzero, at the non-gravitating solutions.

\refout
\end